\documentclass[epj]{svjour}
\usepackage{graphics}

\usepackage{cite}
\usepackage{color}
\usepackage{rotating}
\usepackage{amsmath}
\usepackage{amssymb}
\usepackage{moreverb}
\usepackage[dutch,english]{babel}
\usepackage{esvect}
\usepackage{enumerate}
\usepackage{float}

\newcommand{\pol}[1]{\mathaccent"017E{#1}}
\graphicspath{{Figures/}}
\mathchardef\mhyphen="2D
\begin{document}
\title{Analyzing powers in $d(\pol{p}, pp)n$ at intermediate and large scattering angles at
190~MeV}
\author{M.~Mohammadi-Dadkan\inst{1,2}\thanks{e-mail: m.mohammadi-dadkan@rug.nl} \and H.R.~Amir-Ahmadi\inst{1} \and  M.T.~Bayat\inst{1} \and
	A.~Deltuva\inst{3} \and M.~Eslami-Kalantari\inst{4} \and J.~Golak\inst{5} \and N.~Kalantar-Nayestanaki\inst{1} \and 
	St.~Kistryn\inst{5} \and A.~Kozela\inst{6} \and H.~Mardanpour\inst{1} \and A.A.~Mehmandoost-Khajeh-dad\inst{2}\thanks{e-mail: mehmandoost@phys.usb.ac.ir} \and
	J.G.~Messchendorp\inst{1}\thanks{e-mail: j.g.messchendorp@rug.nl} \and A.~Ramazani-Moghaddam-Arani\inst{7} \and R.~Ramazani-Sharifabadi\inst{1,8} \and
	R.~Skibi{\'n}ski\inst{5} \and E.~Stephan\inst{9} \and H.~Tavakoli-Zaniani\inst{1,4} \and H.~Wita{\l}a\inst{5}
}                     
\offprints{}          
\institute{KVI-CART, University of Groningen, Groningen, The Netherlands
\and
Department of Physics, University of Sistan and Baluchestan, Zahedan, Iran
\and
Institute of Theoretical Physics and Astronomy, Vilnius University, Saul\.{e}tekio al. 3, LT-10222 Vilnius, Lithuania
\and
Department of Physics, School of Science, Yazd University, Yazd, Iran
\and
M.~Smoluchowski Institute of Physics, Jagiellonian University, PL-30348 Krak\'{o}w, Poland
\and
Institute of Nuclear Physics, PAS, PL-31342, Krak\'{o}w, Poland
\and
Department of Physics, Faculty of Science, University of Kashan, Kashan, Iran
\and
Department of Physics, University of Tehran, Tehran, Iran
\and
Institute of Physics, University of Silesia, Chorz\'{o}w, Poland
}
\date{Received: date / Revised version: date}
%
\abstract{
Understanding of the exact nature of three-nucleon forces is the most challenging topic in the field of nuclear physics.
Three-nucleon break-up reaction is a good tool to look into the
underlying dynamics of the nuclear force, 
thanks to its rich kinematical phase space which has different levels of sensitivity 
to three-nucleon force effects.
The recent studies on few-nucleon systems have revealed that the current
nuclear force models cannot describe nucleon-deuteron scattering data
accurately.
In the present work, the analyzing powers of the proton-deuteron break-up reaction
obtained using a $190$~MeV polarized proton beam will be discussed.
We present for the first time the vector analyzing powers for the kinematics in which one of the protons 
scatters to intermediate and large scattering angles at this energy.
The results show a fairly good agreement with various theoretical predictions for
both intermediate and large scattering angles of the break-up phase space.
} 
\authorrunning{M.~Mohammadi-Dadkan }
\titlerunning{Analyzing powers in $d(\vec{p}, pp)n$ at intermediate and large scattering angles...}

\maketitle
\section{Introduction}
\label{intro}
Since 1932 when Chadwick discovered the neutron as a constituent of atomic
nuclei~\cite{Chadwick:1932}, efforts have been made to build a framework for describing
the interaction between nucleons.
Yukawa initiated the first systematic approach for describing the nucleon-nucleon~(NN)
interaction in analogy to the electromagnetic interaction~\cite{Yukawa:1935}.
His formulation of the nuclear force was based on the idea
that the proton and neutron are fundamental particles. Therefore, the potentials
based on these models solely take the nucleons and mesons as degrees of freedom
in the nuclei.
\textcolor{black}{Different NN potentials have been developed in the past decades.
They usually carry the name of the group which developed them such as 
AV18~\cite{Wiringa1995},
CD-Bonn~\cite{Machleidt2001}, etc.. Each of these models
has different parameters and they are fitted to the empirical \textit{pp} and \textit{np} scattering database.}
While these models do a great job in describing two-nucleon systems
below the pion-production threshold energy,
they fail to describe the systems which have more than two nucleons.
There are lots of evidences for this by comparing data and theory for both the nuclear
binding energies~\cite{Pieper2015,Noga2003,Noga2000} and various scattering
observables~\cite{Sakai2000,Ermisch2001,Cadman2001,Hatanaka2002}.   
It has became clear that there are additional underlying dynamics, beyond the NN
interaction, playing a role in the nuclear force and which are referred to as many-body force effects.
We expect that the three-nucleon force (3NF) is the dominant part in the hierarchy of 
many-body force effects.
\begin{figure*}[!t]
\resizebox{0.98\textwidth}{!}{%
  \includegraphics{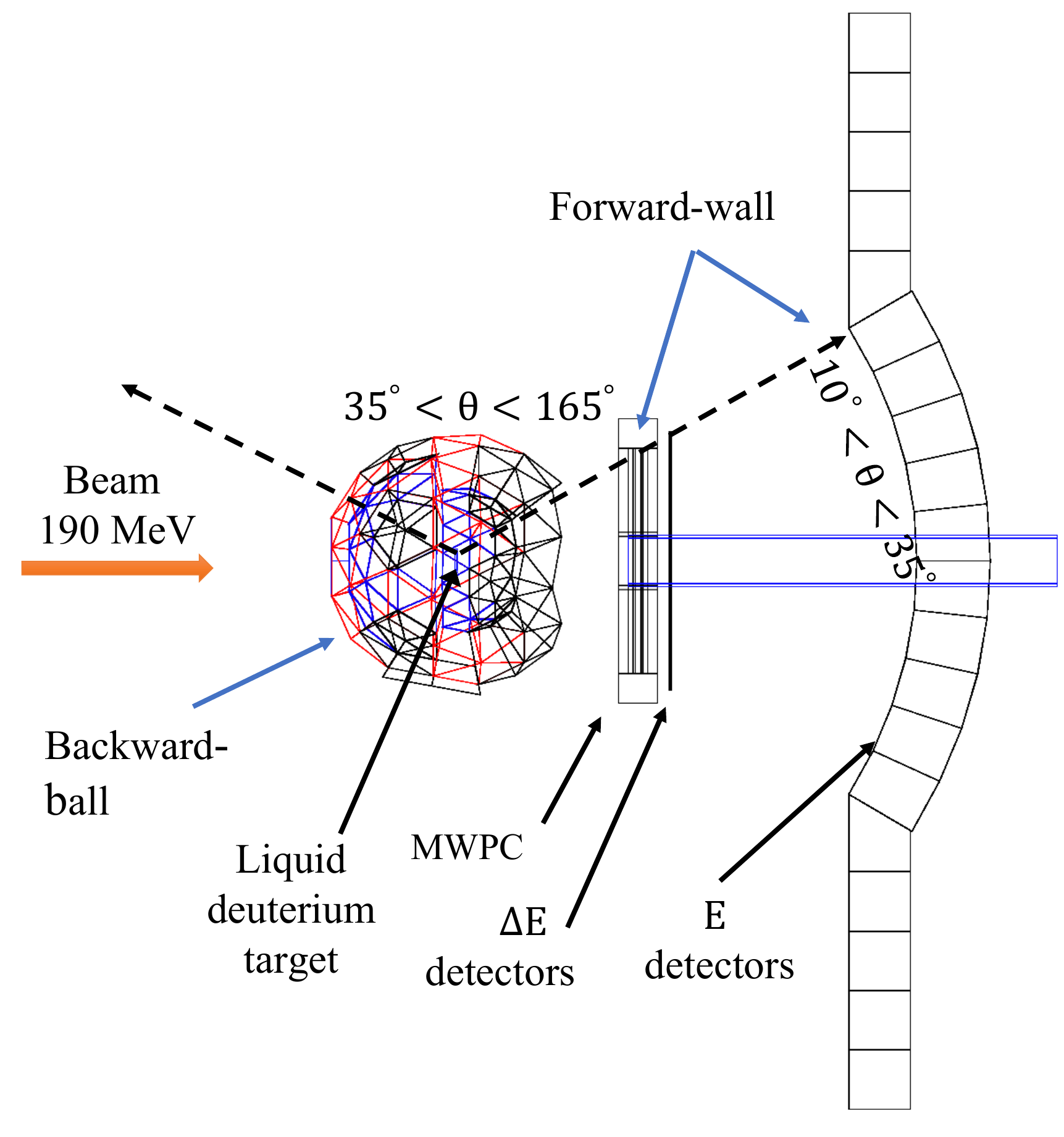}
  \includegraphics{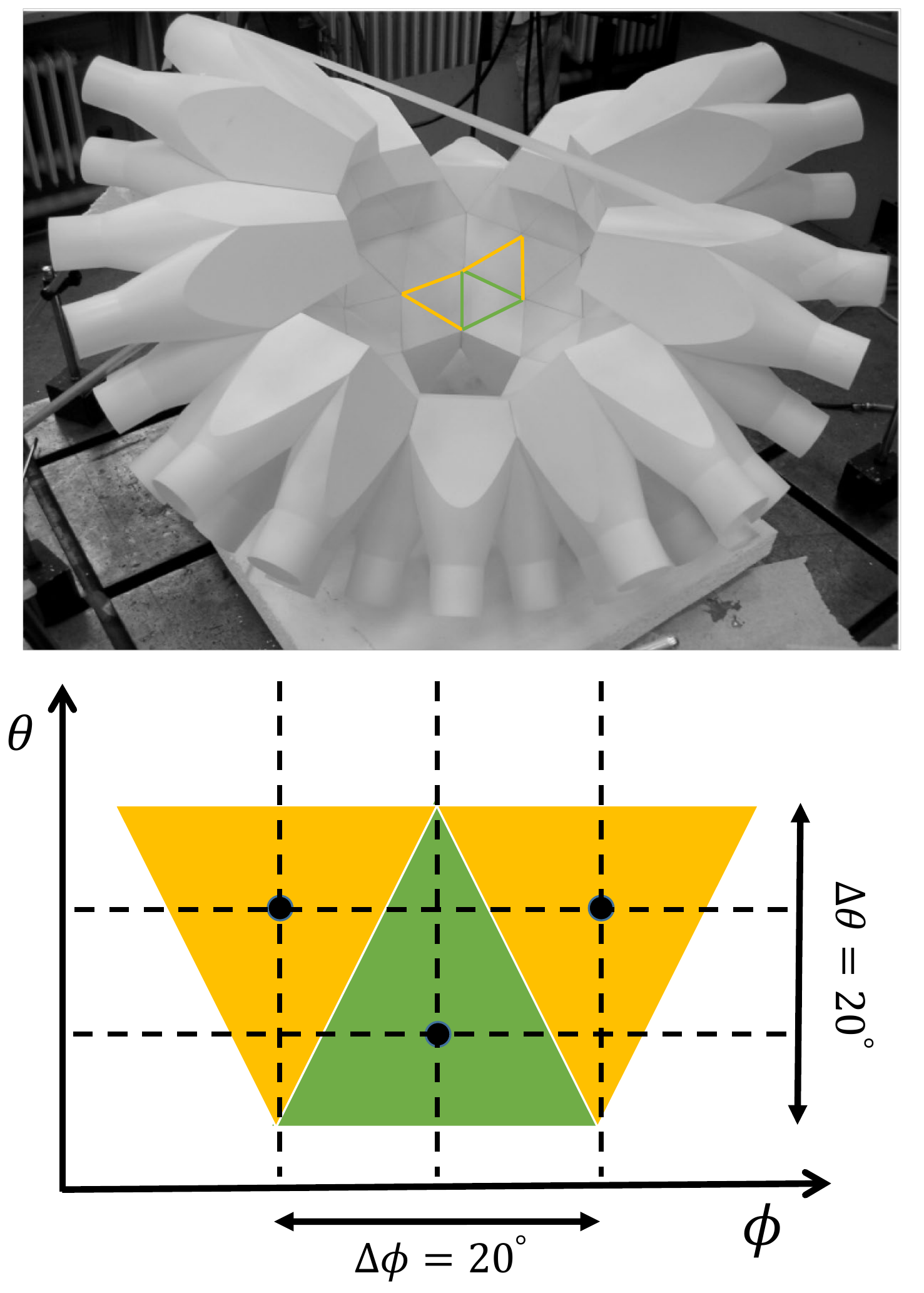}
}
\caption{The left panel illustrates the structure and the components of BINA
(Big Instrument for Nuclear-polarization Analysis) together with
the angular coverage of each part. The right panel shows the partly-assembled
backward-ball (top figure) and the definition of the angles of typical detectors (bottom figure).
The centroid of each ball detector is considered as the angular position of that detector; see the
black points in the bottom-right figure.}
\label{fig:BINA}
\end{figure*}
%
\textcolor{black}{In the past decades, several phenomenological 3NF models such as
Urbana-Illinois (UIX)~\cite{URB1983,URB1997} and CD-Bonn+$\Delta$~\cite{Nemoto1998,Deltuva20032,Deltuva20033}
have been developed based on the works by Fujita and Miyazawa~\cite{FujitaMiyazawa} and Brown and
Green~\cite{BROWN1969} to implement 3NF effects in the interaction of nucleons.}
However, different studies have shown that the inclusion of current models of 3NF do not completely
fill the gap between data and calculations. For a detailed discussion of
the progress of the field at intermediate energies, see Refs.~\cite{Kalantar2011,Kistryn2013}.\par
Ever since it was known that nucleons are composed of quarks and gluons,
the nuclear force has been considered as residual color force between quarks and gluon.
Due to the non-perturbative nature of QCD at low energies, it can, with the present available techniques,
only be applied to the nuclear system through an Effective Field Theory (EFT).
This fundamental approach for nucleonic systems is known as Chiral Perturbation
Theory (ChPT)~\cite{WEINBERG1990}. To take into account all the effective elements in ChPT 
calculations, one should extend the calculations to higher orders.
The work toward higher orders of ChPT is still in
progress. A detailed overview of the present state-of-the-art of the ChPT approach can be found
in Refs.~\cite{Kolck1994,Kolck1999,MACHLEIDT20111,Epelbaum2019,Binder2018}.
\textcolor{black}{Presently, there are no results available from the ChPT calculations for
comparison with the data shown in this paper. For a proton beam energy of $190$~MeV, higher
orders should be included in the calculations before a fair comparison can be made.}\par
In the past decades, the elastic channel in \textit{Nd} scattering has been investigated experimentally at different beam
energies below the pion-production threshold. The results showed that the current 3NF models are not able
to describe the discrepancy between measurements and calculations particularly at higher beam energies and
at the minimum of the differential cross section as well as at backward angles~\cite{Hatanaka2002,Ermisch2005}.
Moreover, the inclusion of 3NFs into 3N Faddeev calculation has failed
to describe the analyzing power $A_{y}$ at low energies, an effect known  in literature as $A_{y}$-puzzle~\cite{AyPuzzle1998,Aypuzzle2001}.
The elastic channel has a limited kinematical phase space and one cannot test all the aspects of the theoretical models.
To have a systematic and a detailed investigation of 3NFs, the three-body break-up channel is a suitable candidate
because of its rich kinematical phase space yielding various degrees of sensitivity to the underlying nucleon-nucleon
and three-nucleon forces.\par
A systematic investigation of 3NFs through elastic and the break-up reaction has been initiated at KVI in a common effort between
the Dutch and the Polish groups since the end of the 90s by developing
various experimental setups and exploiting high-quality polarized
beams~\cite{Ermisch2005,KistrynSALAD2005,KISTRYNCoulomb2006,Stephan2007,Amirahmadi2007,Mehmandoost2011}.  
BINA (Big Instrument for Nuclear-polarization Analysis) is the latest experimental setup which was exploited at KVI
for few-nucleon scattering experiments.
This detection system is capable of measuring the energy and the scattering angles of all the reaction yields of
three and four-body final states in coincidence.
A series of experiments has been conducted using BINA to study the 
deuteron break-up channel using polarized proton and deuteron beams in a wide 
range of energies between 50 and $190$~MeV/nucleon~\cite{Mardanpour_thesis,Eslami_thesis,RamazaniMoghadam_thesis,BINA80mev,BINA100}.
In this article, we present the measurements of the analyzing powers of the proton-deuteron break-up reaction
obtained with a proton-beam energy of $190$~MeV. A specific part of the phase space of this experiment has already been published~\cite{MARDANPOUR2010}.
In this work, we focused on the break-up kinematics in which one proton scatters to forward angles ($\theta_{1}<32^{\circ}$) and 
the other to backward angles ($\theta_{2}>40^{\circ}$).
The experimental setup will be described in Sec.~\ref{sec:setup}. Then, the analysis method and the results will be presented.
\section{Experimental setup}
\label{sec:setup}%
The present experiment was conducted utilizing the AGOR (Acc\'{e}l\'{e}rateur Groningen-ORsay) cyclotron
at KVI delivering a high-quality polarized-proton beam of $190$~MeV.
For details about the experimental setup, see Ref.~\cite{Mardanpour_thesis}.
BINA, is a specially designed 4$\pi$ detection system with which 3NF effects could be studied through
the three-body and four-body elastic as well as break-up reactions.
The detection system consists of different components which make it capable
of measuring energies up to $140$~MeV per nucleon, scattering angles between $10^{\circ}$-$160^{\circ}$, 
and almost a full coverage of azimuthal angles. There are two features that make BINA unique among other
detection systems in the field of few-nucleon scattering:
\begin{itemize}
\item Detection of all the charged particles in the final-state of the three-body and the four-body reactions in coincidence;
\item The almost 4$\pi$ angular coverage which can probe a large part of the phase space of the break-up reactions.
\end{itemize}
In general, BINA has two main parts, the forward-wall and the backward-ball.
The left part of Fig.~\ref{fig:BINA} illustrates different parts of BINA.
The forward-wall consists of a cylindrically-shaped array of scintillators ($E$-detectors) to measure the energy of
charged particles, a MWPC to measure their scattering angles, and a vertical array of $\Delta E$-detectors that is
used in combination with $E$-detectors for particle identification. 
The forward-wall covers angles which are in the range of $10^{\circ}<\theta<35^{\circ}$.
The backward-ball is a ball-shape scattering chamber which consists of 149 pyramid-shaped plastic scintillators.
The geometrical design of the backward-ball and its building blocks are similar to the classic soccer ball.
The position of each backward-ball scintillator is labeled with a specific $\theta$ and $\phi$ in the laboratory coordinate
system. The backward-ball can measure the scattering angles of the particles with a precision of
$\pm10^{\circ}$. The right panels of Fig.~\ref{fig:BINA} demonstrate the structure of the backward-ball and
the positioning of the detector elements.
This part of BINA is capable of measuring the energy and the scattering angles of the particles in the range
of $35^{\circ}<\theta<160^{\circ}$ except where the target holder is attached to the scattering chamber.
Therefore, there are no detectors at the part of the phase space with polar and azimuthal angles in the range of
$80^{\circ}<\theta<120^{\circ}$ and $70^{\circ}<\phi<110^{\circ}$, respectively. Also, at the angular range of
$80^{\circ}<\theta<100^{\circ}$, the scattered protons partially lose their energies in the target-holder ring.
Due to the complicated procedure for the energy correction for these protons, we decided to leave out these detectors from the analysis.
The detectors which are located at $\theta>140^{\circ}$ are excluded from this analysis as well because the expected energy of the break-up
protons is lower than the threshold of these detectors.
The parts that were used in the analysis cover scattering angles within $40^{\circ}<\theta<80^{\circ}$ and $100^{\circ}<\theta<140^{\circ}$.
In the present analysis, we obtained the analyzing powers of the proton-deuteron break-up reaction for the kinematics in which one
proton was registered in the forward-wall and the other one in the backward-ball.

\section{Analysis method}
\label{sec:analysis}
\begin{figure}[t]
\resizebox{0.49\textwidth}{!}{%
  \includegraphics{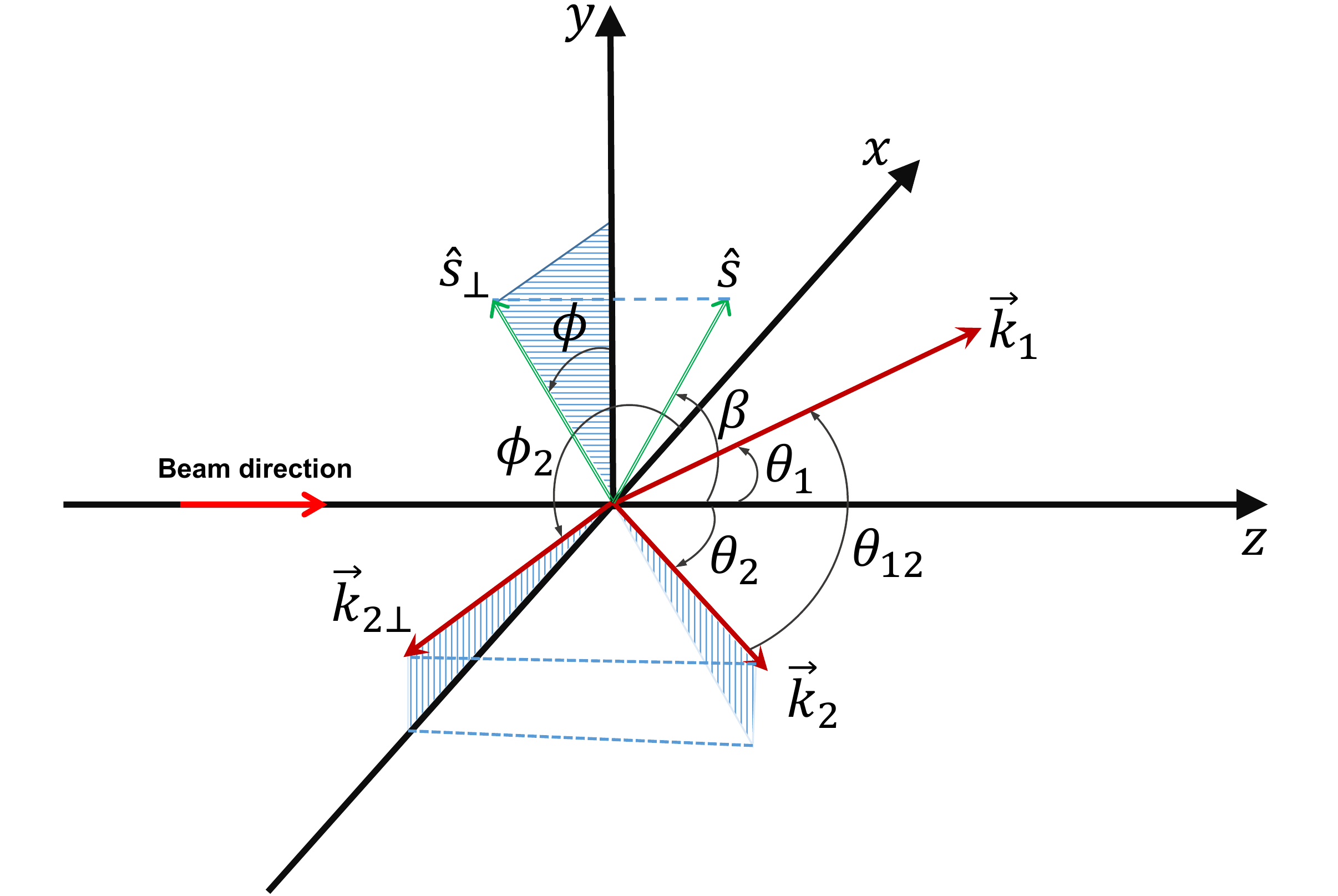}
}
\caption{The scattering diagram of a three-body final-state reaction. 
$\pol{k}$$_1$ and $\pol{k}_2$ are the momentum of the two detected particles with
scattering angels $\theta_{1}$ and $\theta_{2}$, respectively. $\phi_{2}$ is the
angle between the projection of $\pol{k}_2$ on the $x-y$ plane and positive
direction of the $x$-axis.
$\hat{s}$ denotes the direction of spin of the projectile.
$\phi$ is the angle between the projection of $\hat{s}$ on the $x-y$ plane
and the $y$-axis.
$\beta$ is the angle between spin of the projectile and its momentum
direction ($z$-axis). In the present experiment, the polarization of beam
is perpendicular to the direction of the beam momentum and, therefore $\beta=90^{\circ}$.
}
\label{fig:scattering}       
\end{figure}

\begin{figure*}[!h]
\resizebox{0.9\textwidth}{!}{%
  \includegraphics{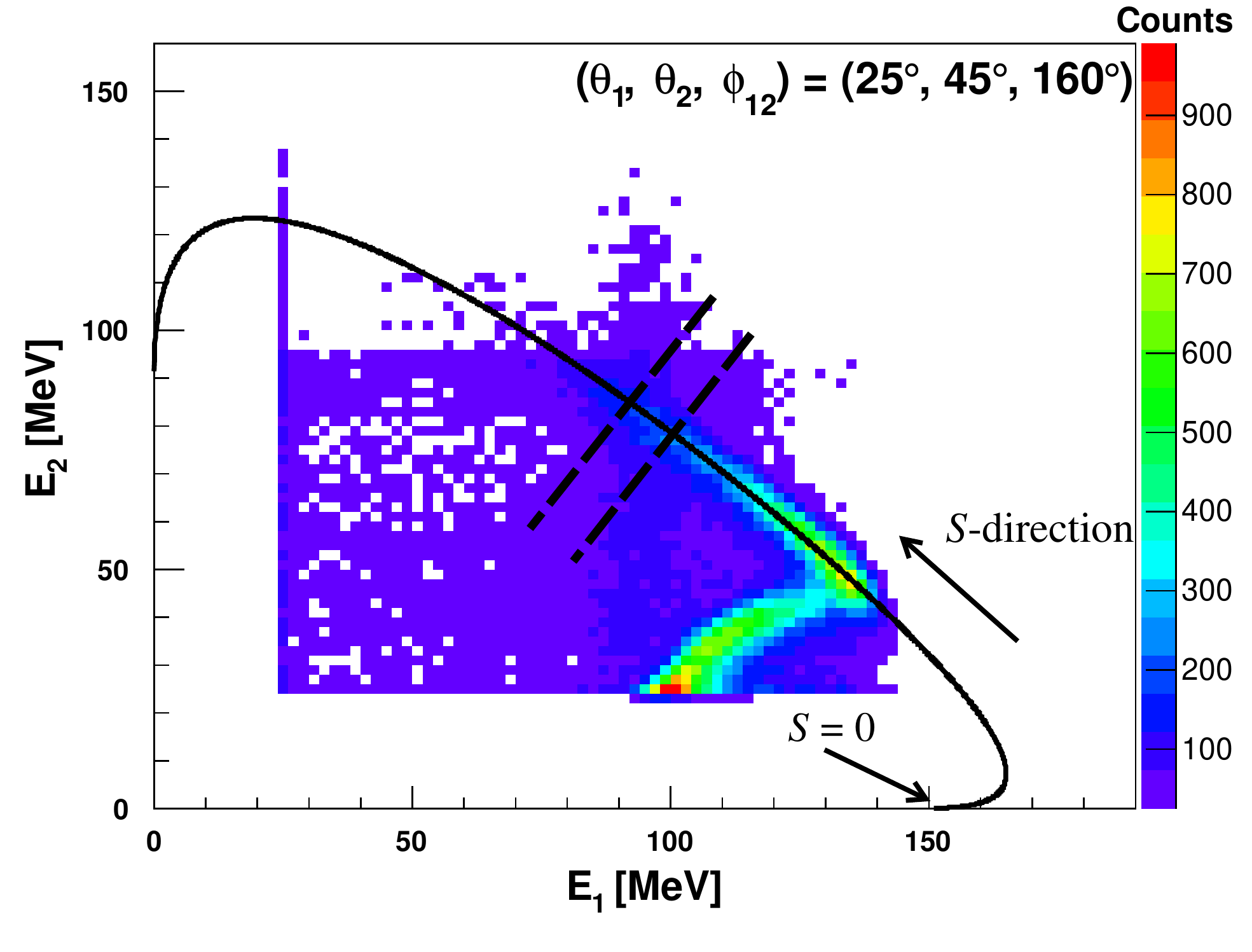}
}
\caption{The kinematically allowed relation of the energies of the two protons of the three-body
break-up reaction, so called $S$-curve, for
($\theta_{1}=25^{\circ}, \theta_{2}=45^{\circ}, \phi_{12}=160^{\circ}$) along with the experimental data.
In this analysis, we took the minimum of $E_{2}$ as the starting point of $S$-value.
The protons which have energies $E_{1}>140$~MeV punch through the E-detectors
of the forward-wall and deposit part of their energies in the detectors resulting in a kink in the spectrum.
\textcolor{black}{The dashed lines illustrate the definition of a typical $S$-bin for $S=127$~MeV.}
The events within each $S$-bin were used to extract the asymmetry distributions and the analyzing powers.
}
\label{fig:scurve}       
\end{figure*}

In the proton-deuteron break-up reaction, there are three particles in the final state, two protons
and one neutron. From the kinematic point of view, there are 9 variables
$(\theta_{i},\phi_{i}, E_{i})$ for this reaction channel which can be measured.
Applying energy and momenta conservation reduces the number of degrees of freedom to 5.
It is, therefore, sufficient to measure 5 of the 9 variables to have all the information of the reaction.
BINA is able to measure the energy and the scattering angles of two protons in coincidence which 
provides an extra redundancy of one degree of freedom.
Conventionally, in the three-body break-up reaction, each kinematical configuration is defined with
$(\theta_{1}, \theta_{2}, \phi_{12}=|\phi_{2}-\phi_{1}|)$ which represents the polar scattering angles
of protons 1 and 2 and the relative azimuthal angle between them.
Figure~\ref{fig:scattering} shows the definition of the scattering angles of two protons
in the $p$-$d$ break-up reaction. There are two conventions for defining the $y$-axis~\cite{OHLSEN1981}.
In this paper, we use the asymmetric choice for the azimuthal angles, where $\vec{k}_{1}$ lies in the
$xz$ plane, and therefore $\phi_{1}=0$, see Fig.~\ref{fig:scattering}.
\textcolor{black}{Here, $\vec{k}_{1}$ is the momentum of the outgoing proton which scattered
to smaller polar angle $\theta$.}
For each kinematical configuration, the energy correlation of the two protons is referred to as the $S$-curve.
By employing the $S$-curve, a variable $S$ is defined as an arc length of the curve with a
starting point at the minimum energy of one of the protons.
In the present work, we used a convention in which the starting point of $S$ is the minimum value of 
$E_{2}$ and it increases counter-clockwise as it is shown in Fig.~\ref{fig:scurve}.
\textcolor{black}{Here, $E_{1}$ and $E_{2}$ correspond to $\vec{k}_{1}$ and $\vec{k}_{2}$, respectively.}
Traditionally, the three-body break-up observables are presented as a function of $S$-value.\par
\begin{figure}[!t]
\resizebox{0.49\textwidth}{!}{%
  \includegraphics{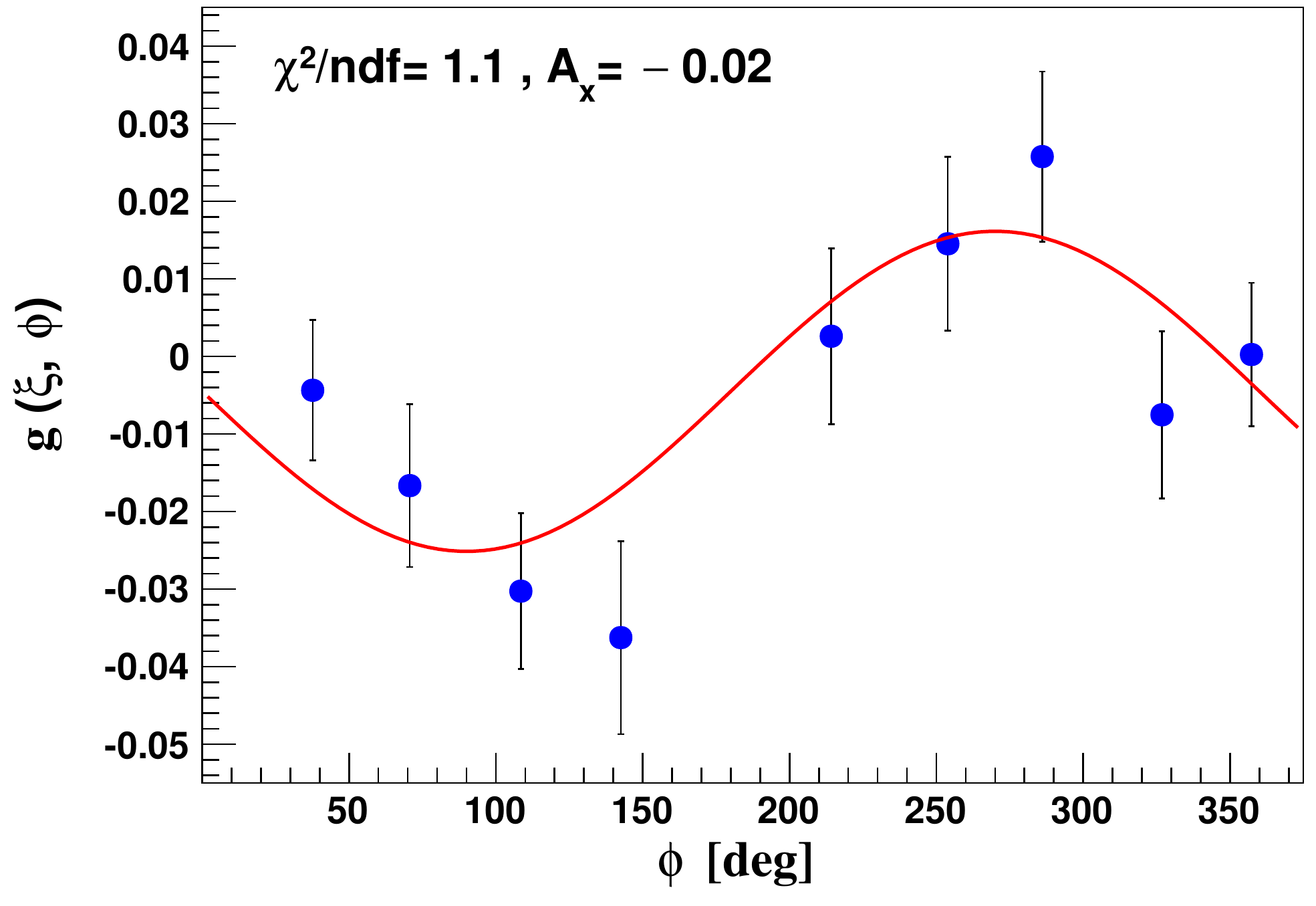}
}
\resizebox{0.49\textwidth}{!}{%
  \includegraphics{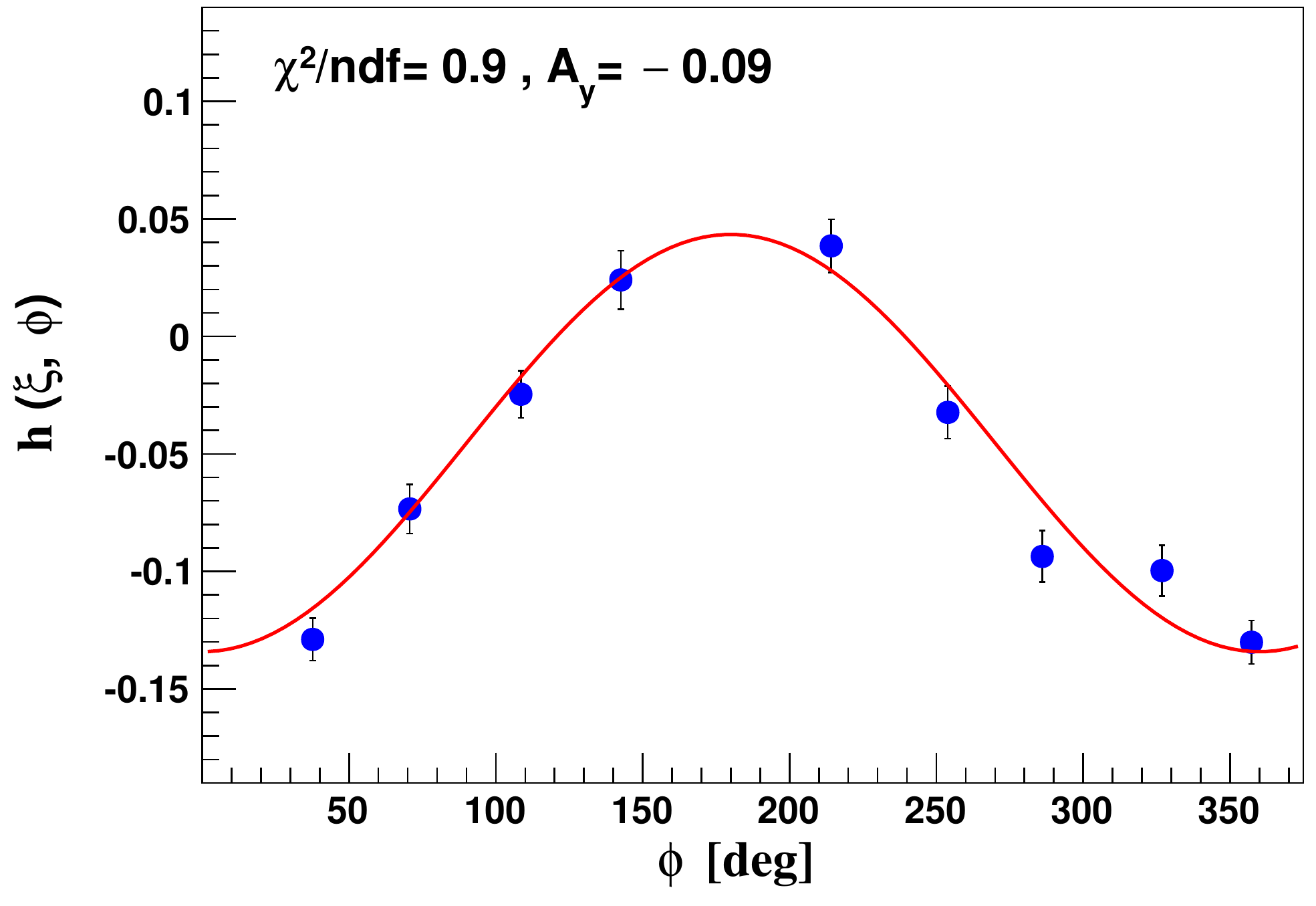}
}
\caption{The constructed asymmetry distributions for $A_{x}$ (top panel) and
$A_{y}$ (bottom panel) for the kinematics ($25^{\circ},45^{\circ}, 160^{\circ}$) and $S=76$~MeV.
The quality of fit and the obtained values of the analyzing powers are shown
at the top left of each panel.
The statistical errors are obtained using the number of counts before subtracting
the background. 
}
\label{fig:assym}       
\end{figure}

To investigate spin effects in the nuclear force, one should measure various spin-related observables
such as analyzing powers. The polarized beam (or target) imposes an asymmetry on the cross section of the
break-up reaction. In this article, we present a study of the vector analyzing powers, $A_{x}$ and $A_{y}$,
in the proton-deuteron break-up reaction. $A_{x}$ is odd and $A_{y}$ is even under the parity inversion.
Here, the parity inversion corresponds to changing the sign of the azimuthal scattering angle of
the two outgoing protons.
By exploiting the formulation taken from Ref.~\cite{OHLSEN1981} for spin-$\frac{1}{2}$ protons,
the following expressions can be derived for two parity modes:
\begin{equation}
 \Upsilon^{+}(\xi,\phi)=\frac{N_{p}^{\uparrow}-N_{p}^{\downarrow}}{N_{p}^{\downarrow}p_{z}^{\uparrow}-N_{p}^{\uparrow}p_{z}^{\downarrow}}= A_{y}\cos\phi-A_{x}\sin\phi,
 \label{eq:Assymetry-p}
\end{equation} 
\begin{equation}
 \Upsilon^{-}(\xi,\phi)=\frac{N_{m}^{\uparrow}-N_{m}^{\downarrow}}{N_{m}^{\downarrow}p_{z}^{\uparrow}-N_{m}^{\uparrow}p_{z}^{\downarrow}}= A_{y}\cos\phi+A_{x}\sin\phi.
 \label{eq:Assymetry-m}
\end{equation}
where $\Upsilon^{+}(\xi,\phi)$ and $\Upsilon^{-}(\xi,\phi)$ are the asymmetry terms for two parity modes
and $\xi$ represents any appropriate set of kinematical variables.
$N_{p}^{\uparrow}$ ($N_{p}^{\downarrow}$) are the number of events for up (down) polarization modes of
the ($\phi_{1},\phi_{2}$) setup. Similarly, $N_{m}^{\uparrow}$ ($N_{m}^{\downarrow}$) are the number of events for up
(down) polarization modes of the ($-\phi_{1},-\phi_{2}$) setup.
$p_{z}^{\uparrow}$ and $p_{z}^{\downarrow}$ are the polarization degrees of up and down modes, respectively.
$A_{x}$ and $A_{y}$ are the two components of the vector analyzing power and $\phi$ is the azimuthal angle of the reaction plane.
Two asymmetry components can be constructed from $\Upsilon^{+}(\phi)$ and $\Upsilon^{-}(\phi)$ to
extract $A_{x}$ and $A_{y}$ independently:
\begin{equation}
g(\xi,\phi)=\frac{\Upsilon^{-}(\xi,\phi)-\Upsilon^{+}(\xi,\phi)}{2}=A_{x}\sin\phi,
\label{eq:APx}
\end{equation} 
\begin{equation}
h(\xi,\phi)=\frac{\Upsilon^{-}(\xi,\phi)+\Upsilon^{+}(\xi,\phi)}{2}=A_{y}\cos\phi.
\label{eq:APy}
\end{equation}
\begin{figure*}[t]
\resizebox{0.9\textwidth}{!}{%
  \includegraphics{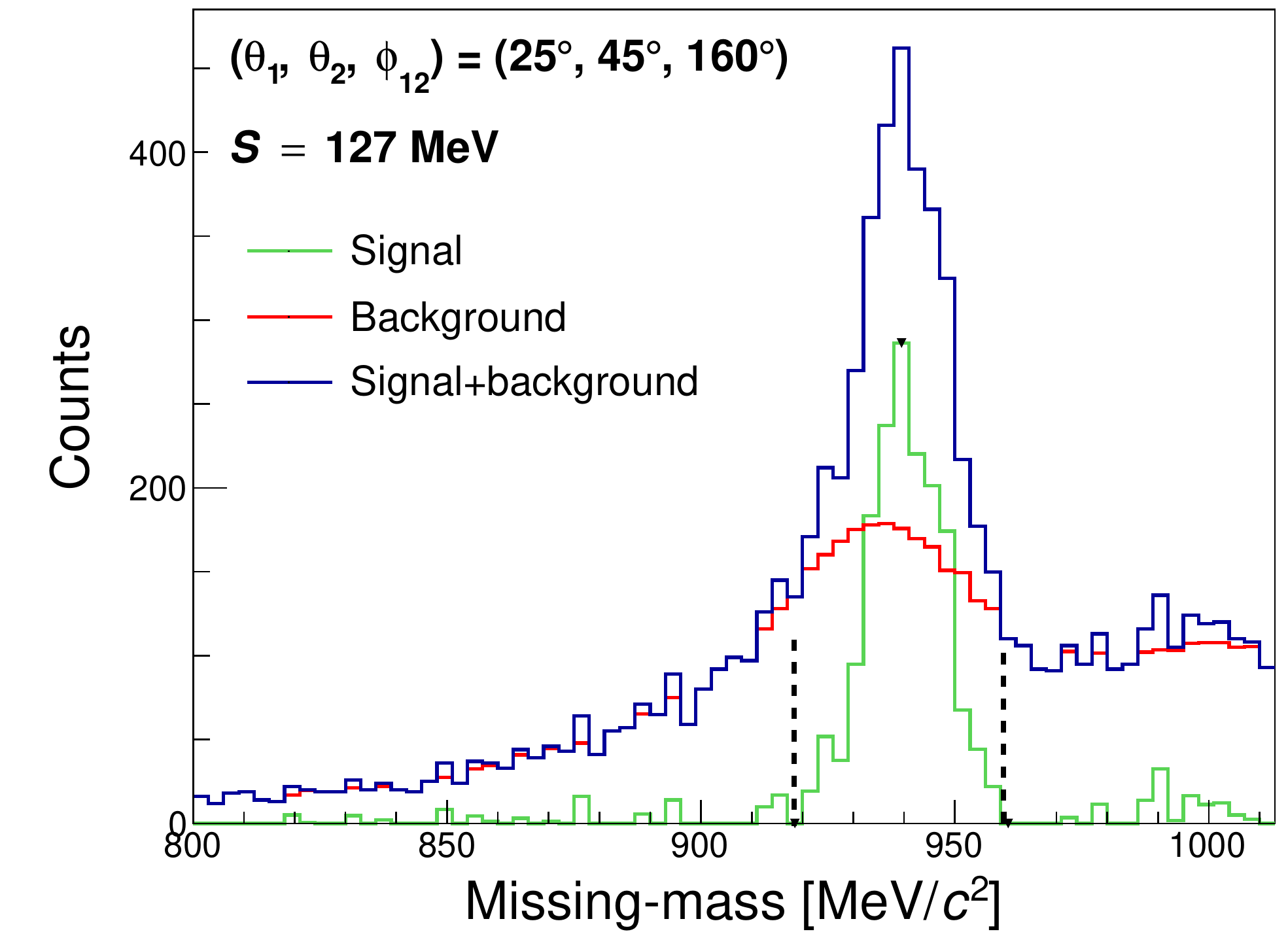}
}
\caption{The reconstructed neutron missing-mass for one of the $S$-bins.
The peak position of the spectrum which is around the expected 
missing-mass of neutron indicates that the detectors are well calibrated.
The interval of $\pm3\sigma$ was used to obtain the number of break-up events under the
peak after subtracting the background.
The background was estimated using the sensitive
non-linear iterative peak clipping algorithm (red line). The green histogram
shows the missing-mass spectrum after background subtraction.
The vertical black dash lines illustrate the integration window.
}
\label{fig:missmass}       
\end{figure*}
After the energy calibration of the forward-wall detectors~\cite{Mardanpour_thesis}, each of the
backward-ball detectors was calibrated using the energy information of a break-up proton~(1)
registered in the forward-wall in coincidence with the other proton~(2) arriving at one of the
backward-ball detectors. Further, the protons with energies larger than
$140$~MeV ($E_{1}>140$~MeV) punch through the $E$-detectors of the forward-wall and consequently deposit
part of their energies in the detectors, see Fig.~\ref{fig:scurve}.
The points corresponding to the region of punch through are excluded
in the final spectrum.
Events were selected for each kinematic
of ($\theta_{1}\pm 2.5^{\circ}, \theta_{2}\pm 10^{\circ}, \phi_{12}\pm 10^{\circ} $).
The $S$-curve is divided into 15 equally-sized bins and the events within each $S$-bin are used to reconstruct
the missing-mass of the neutron.
Figure~\ref{fig:scurve} shows the selected events for configuration of
$(\theta_{1}, \theta_{2}, \phi_{12})=(25^{\circ}, 45^{\circ}, 160^{\circ})$
together with the corresponding $S$-curve.
The number of counts was obtained for each spin state and parity mode using the corresponding
missing-mass spectrum after subtracting the background, see Fig.~\ref{fig:missmass}.
The background mainly stems from uncorrelated protons from two different reactions, either elastically
scattered or from a break-up reaction, and accepted by the hardware trigger as candidate break-up events.
\textcolor{black}{Figure~\ref{fig:missmass} shows the missing-mass
spectrum for the $S$-bin shown in Fig.~\ref{fig:scurve}.
The cut on the $S$-curve induces an artificial correlation between the two time-uncorrelated
hits that mainly stem from two protons of the elastic proton-deuteron scattering process and
of the break-up reaction. As a consequence, a peaking structure of the background is observed
in the missing-mass spectrum for the selected $S$-bin. The location of this peak depends upon
$S$ and is well understood.}
To account for the background including the one from the elastic channel,
we used an algorithm which was developed to estimate background of
gamma-rays spectrum~\cite{MORHAC1997}. This method allows to separate continuous background from peaks,
based on a sensitive non-linear iterative peak clipping algorithm.
The algorithm was applied to each neutron missing-mass
spectrum to estimate the background shape using an $8^{\rm th}$-order polynomial function.
The main motivation for employing this method was the complicated shape of
the background for each detector and spin state.
Figure~\ref{fig:missmass} shows the estimated background shape for a
typical missing-mass spectrum.
To find the peak position of the missing-mass spectrum, we used an algorithm
for peak searching~\cite{MORHAC2000}. It allows to automatically identify the
peaks in a spectrum in the presence of the continuous background and statistical
fluctuations. The algorithm is based on smoothed second differences that are compared
to its standard deviations. Therefore, it is necessary to pass a value for the width of
the peak ($\sigma$) to the peak searching function.
The algorithm is selective to the peaks with a given $\sigma$.
Here, we fixed $\sigma$ to $7$~MeV/$c^{2}$ which is equal to the expected width of the peak in the
missing-mass spectrum.
Also, to have a consistent counting of events in the four missing-mass spectra, spin up and down and two parity
modes, the peak position of one of the spectra was used for the other spectra.
The number of counts was obtained by including the events which were located $\pm3\sigma$
of the peak position.
To extract the analyzing powers from Eqs.~\ref{eq:APx} and \ref{eq:APy}, the number of counts for each polarization
and parity modes are extracted and normalized with the integrated charge from the Faraday cup. The $g(\xi,\phi)$ and 
$h(\xi,\phi)$ are constructed using the normalized counts as a function of $\phi$.
To measure the values of analyzing powers, the functions $A\sin\phi+B$ and
$C\cos\phi+D$ are fitted to $g(\xi,\phi)$ and $h(\xi,\phi)$, respectively. The parameters $A$ and $C$
represent the analyzing powers $A_{x}$ and $A_{y}$, respectively. The parameters $B$ and $D$ are free 
parameters to correct for a possible offset of the charge measurement and other normalization factors which we missed during the analysis.
Figure~\ref{fig:assym} shows the results of the fits for $g(\xi,\phi)$ and $h(\xi,\phi)$ for one of the analyzed configurations.
The quality of the fits for all configurations has been found to be good and an average
$\chi^{2}/$ndf$~\sim 1$ was obtained.
\section{Results and discussion}
\label{sec:result}
\begin{figure*}[!ht]
\centering
\resizebox{0.85\textwidth}{!}{
 \includegraphics{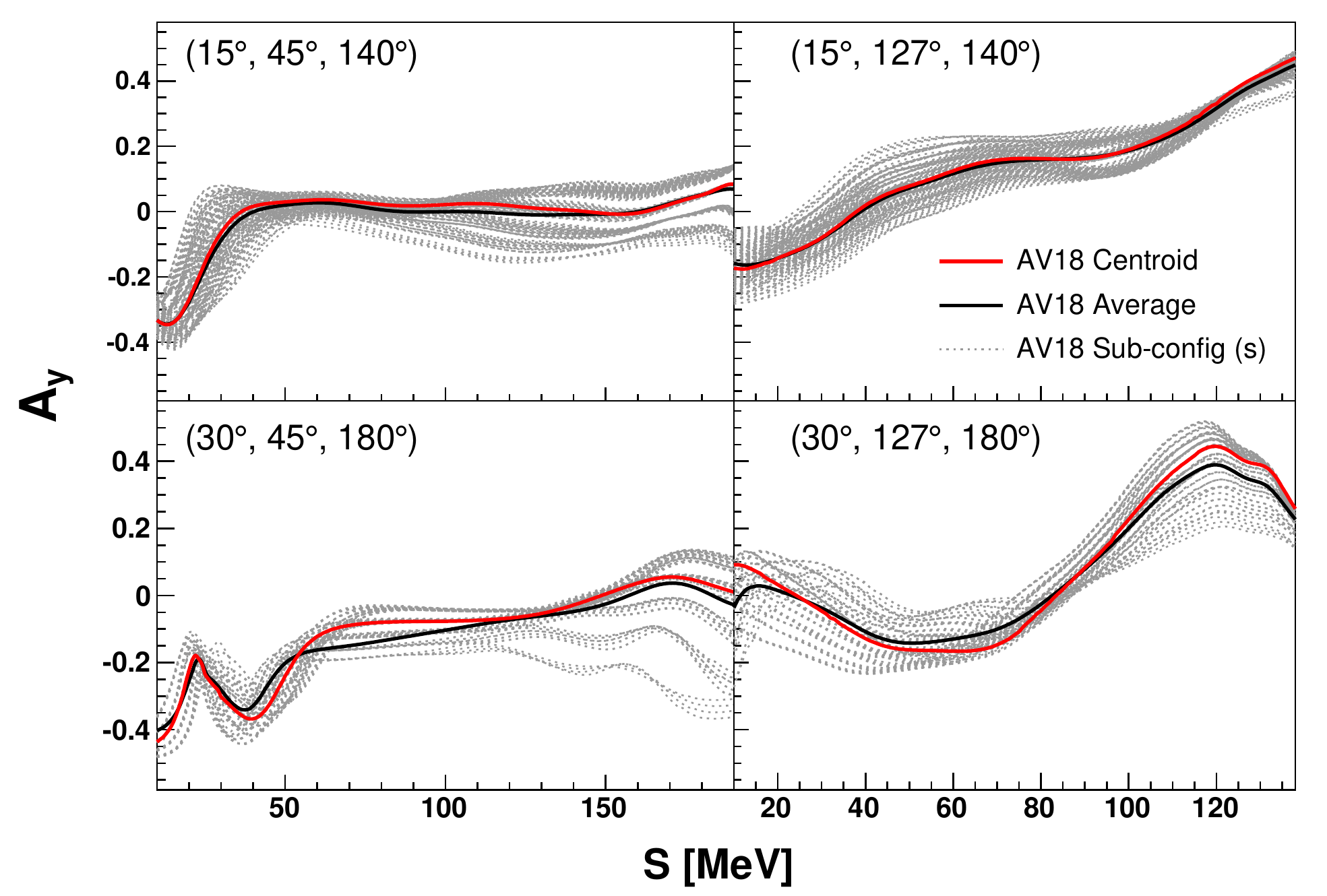}
}
\caption{Comparison of the theoretical predictions (AV18) at the center of the
angular bins (red line) and averaged over the angular bins (black line) for
some kinematical configurations. The gray lines are the sub-configurations
which are within the detector acceptance and are used to perform averaging.}
\label{fig:Average}       
\end{figure*}

The proton-deuteron break-up reaction has a rich kinematical phase space. In the present work, we measured the vector
analyzing powers for the kinematics in which one proton scatters to forward angles ($\theta_{1}<32^{\circ}$)
and the other to intermediate ($\theta_{2}=45^{\circ}$) and large ($\theta_{2}=134^{\circ}$)
scattering angles.
Due to the low angular resolution in the backward ball, all the theoretical calculations
have been averaged over the experimental angular bins. For the averaging, we used the theoretical cross sections and
solid angles as weighting factors. For each non-coplanar (coplanar) kinematic, we exploited 125 (75) sub-configurations
with a $1^{\circ}$ step in $\theta_{1}$, $4^{\circ}$ step in $\theta_{2}$, and $2^{\circ}$ step in $\phi_{12}$.
Figure~\ref{fig:Average} shows a comparison between the results of the averaging and the central values for some
configurations.
Even though the averaging does not show a significant deviation from the calculation based
on the central value of the angle of the bin for most kinematics, we performed this procedure for all of them
to have a consistent comparison between data and theory.

\begin{figure*}[!b]
\resizebox{0.99\textwidth}{!}{
 \includegraphics{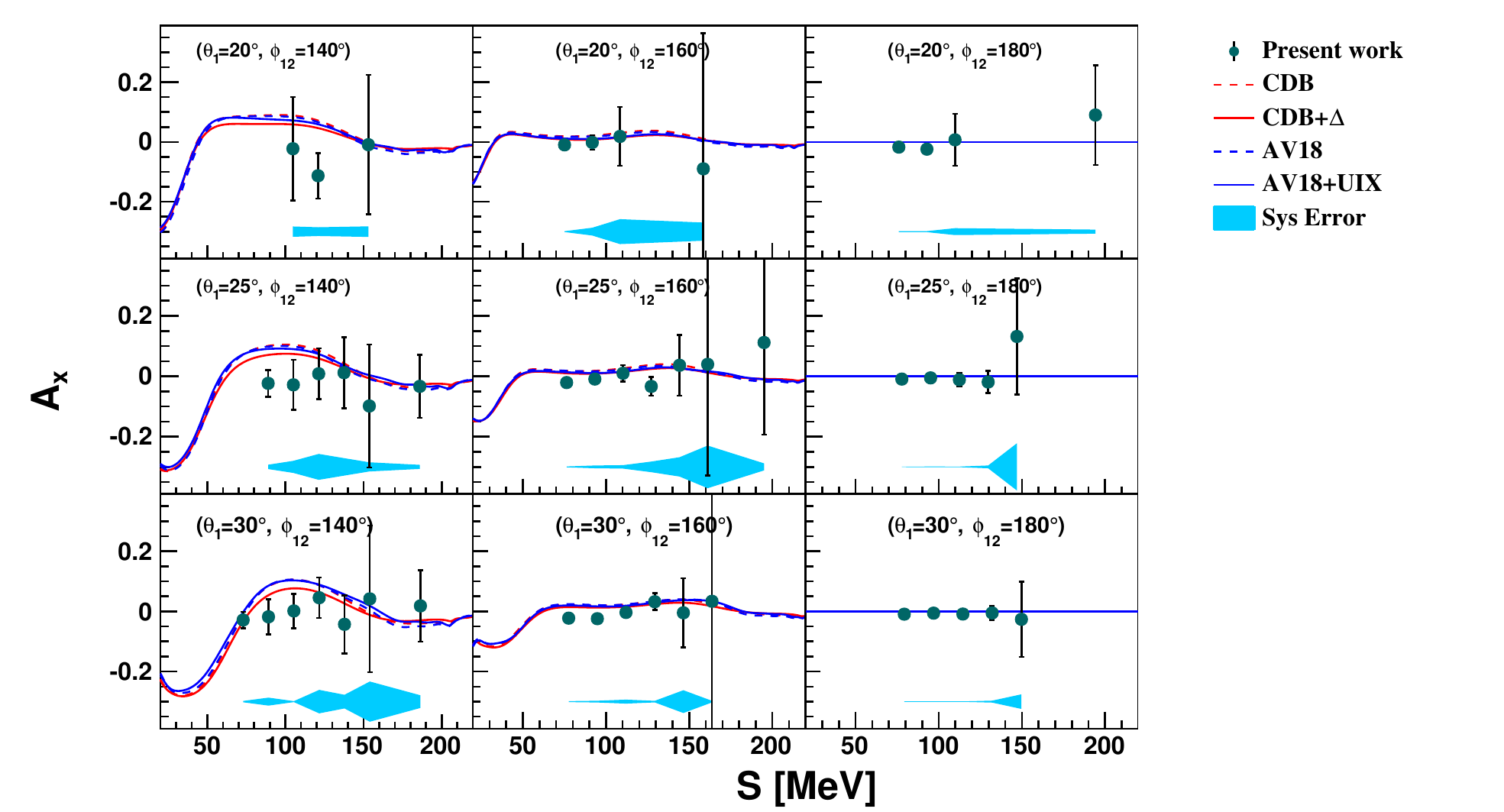}
}
\caption{Experimental vector analyzing power $A_{x}$ for $\theta_{2}=45^{\circ}$
and for various combinations of ($\theta_{1}, \phi_{12}$), as indicated in each panel.
Data are compared with predictions of various theoretical approaches based on pairwise NN interactions
alone NN (dash lines) as well as with 3NF included NN+3NF (solid lines); see the legend.
The error bars represent statistical uncertainties.
The cyan band depicts the systematic uncertainty ($2\sigma$) which stems from different sources; see the text.}
\label{fig:Ax_45}       
\end{figure*}

\begin{figure*}[!b]
\resizebox{0.99\textwidth}{!}{
 \includegraphics{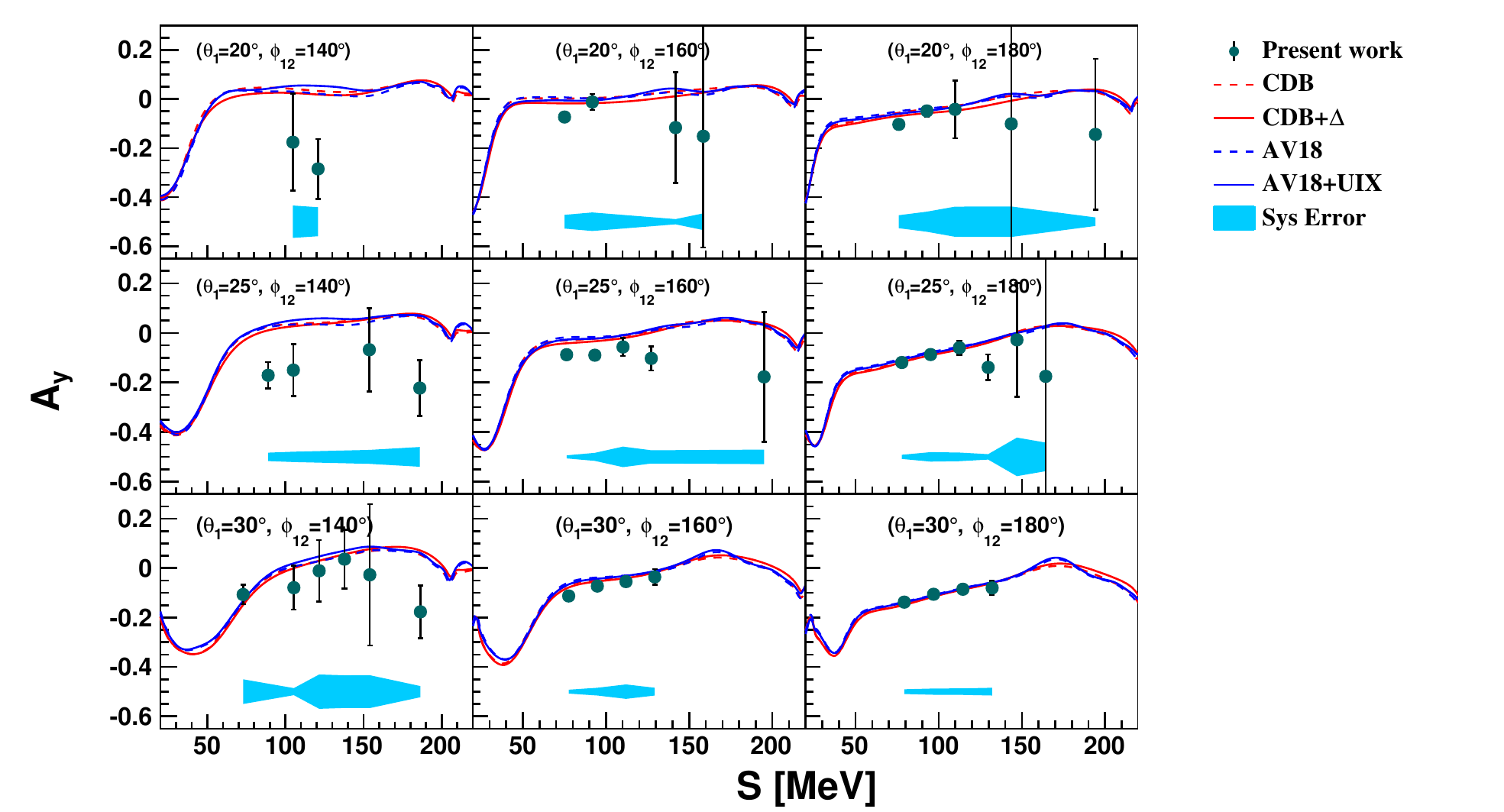}
}
\caption{Same as Figure~\ref{fig:Ax_45} but for $A_{y}$ instead of $A_{x}$.}
\label{fig:Ay_45}       
\end{figure*}

\begin{figure*}[!b]
\resizebox{0.99\textwidth}{!}{
 \includegraphics{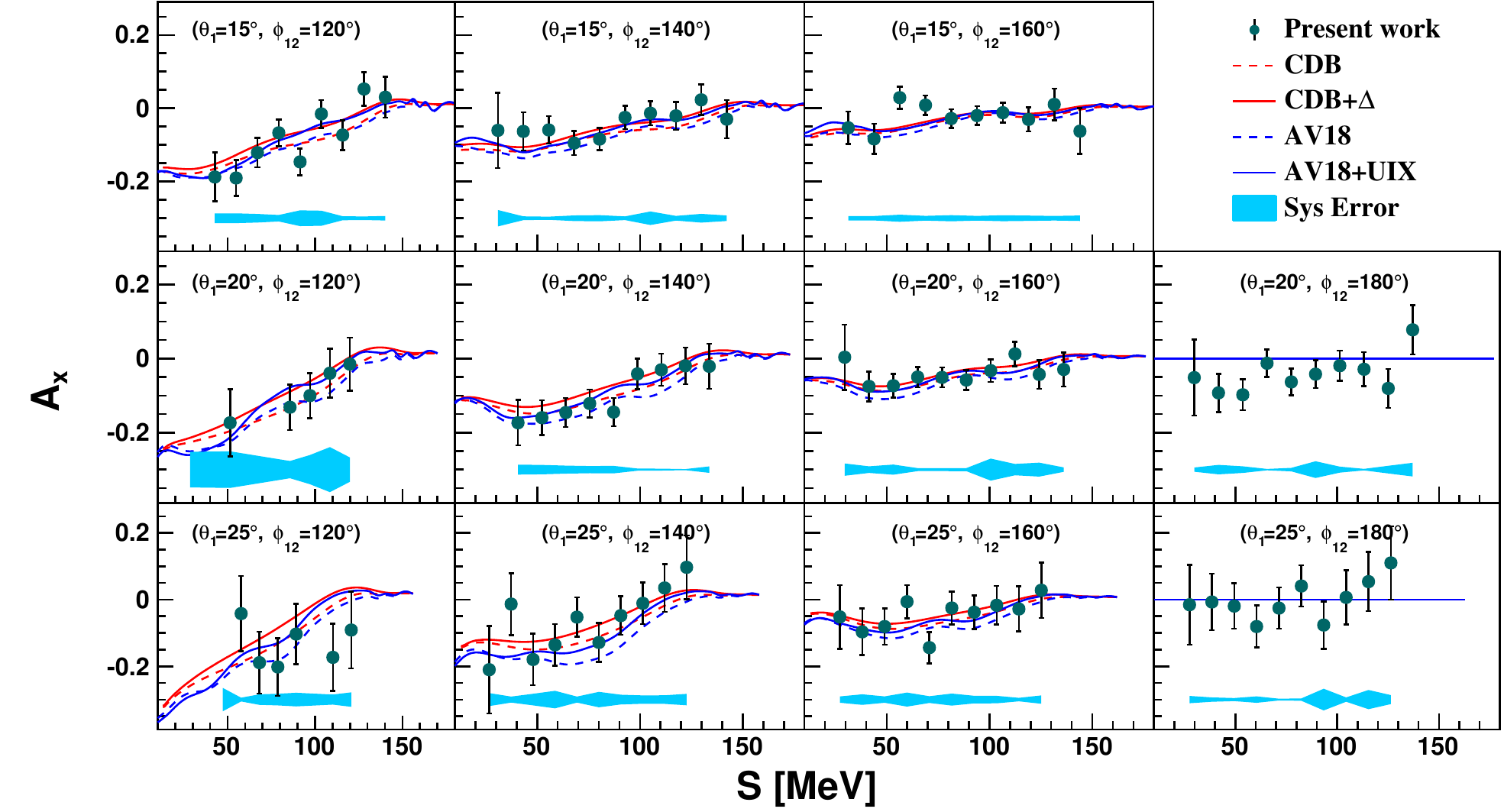}
}
\caption{Same as Figure~\ref{fig:Ax_45} but for $\theta_{2}=134^{\circ}$.}
\label{fig:Ax_134}       
\end{figure*}

\begin{figure*}[!b]
\resizebox{0.99\textwidth}{!}{
 \includegraphics{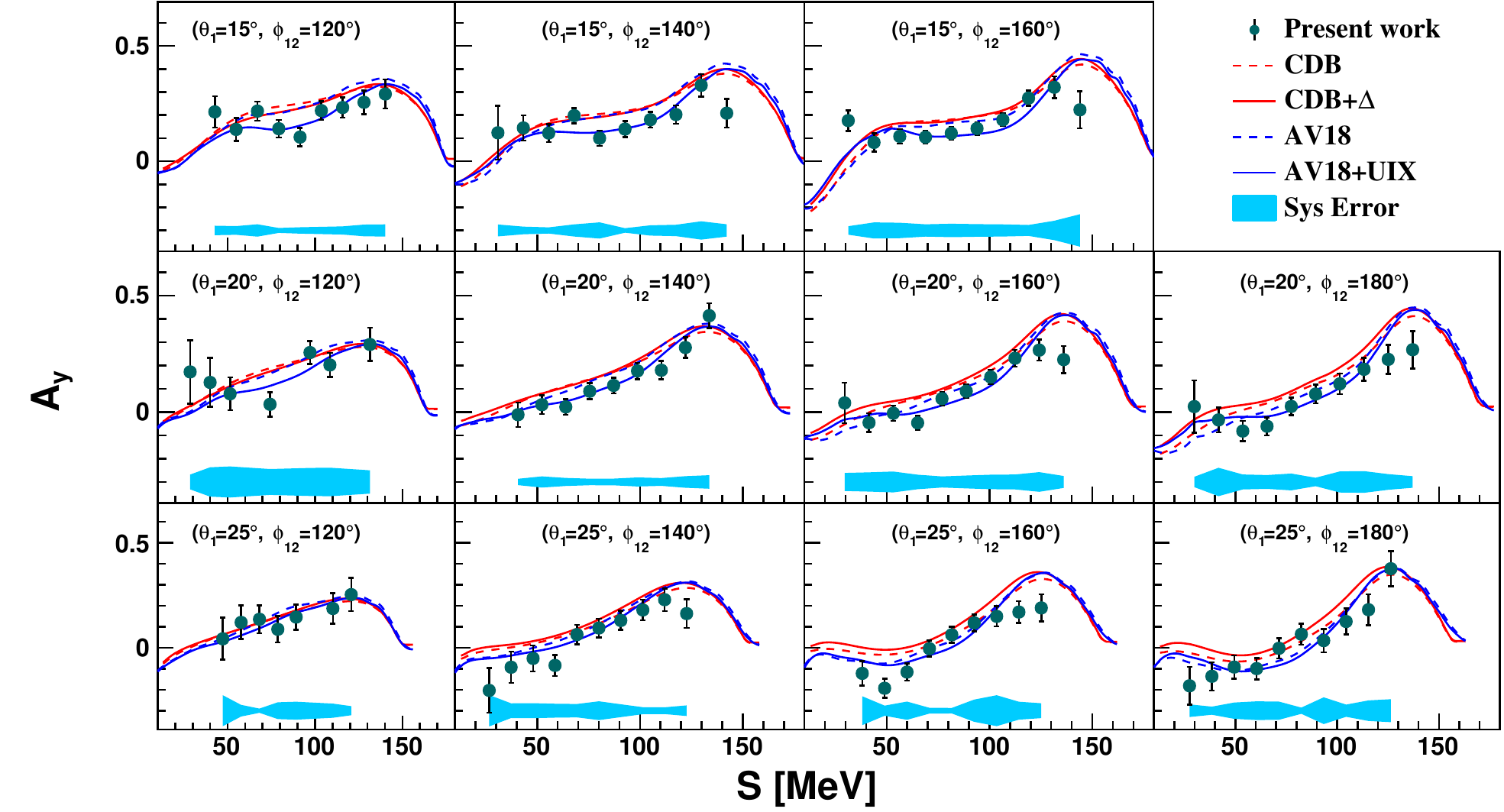}
}
\caption{Same as Figure~\ref{fig:Ax_134} but for $A_{y}$ instead of $A_{x}$.}
\label{fig:Ay_134}       
\end{figure*}

The measured analyzing powers are presented in
Figs.~\ref{fig:Ax_45}-\ref{fig:Ay_134} for a set of configurations
at intermediate ($\theta_{2}=45^{\circ}$) and large ($\theta_{2}=134^{\circ}$) scattering angles,
and compared with Faddeev calculations using various NN and NN+3NF
potentials. The red dash and solid lines represent the predictions of CD-Bonn and 
CD-Bonn+$\Delta$~\cite{Deltuva20032,Deltuva20031,Deltuva20033}, respectively.
The blue dash and solid lines are the predictions of AV18 and AV18+UIX~\cite{GLOCKLE1996,Witala2002,WiatalaComp2002},
respectively. The results of all the calculations are averaged over the experimental angular bins.
\textcolor{black}{The Coulomb effect is negligible in the analyzing powers~\cite{MARDANPOUR2010}. Therefore, the results of
CD-Bonn+$\Delta$+Coulomb are not included for ease of comparison with the other calculation.}
The cyan band in each figure represents the systematic uncertainty which consists of three
parts. The polarization of the beam was measured by analyzing the elastic proton-deuteron channel with BINA and by
using the measurements of the In-Beam Polarimeter (IBP).
Typically, the beam polarization varied gradually between $0.5$-$0.7$ during the course
of the experiment.
The associated uncertainties of the polarization values are $6\%$ statistical and $3\%$ systematical~\cite{Mardanpour_thesis}.
This uncertainty in the polarization causes a systematic error of $\sigma_{pol}\sim 7\%$ in the analyzing
power measurement.
Another source of systematic uncertainty is identified to be from residual and unknown
asymmetries caused by efficiency variations between the up and down polarization states
of the beam, error in charge normalization, etc.. All of these will result in a “wrong” model presented
in Eqs.~\ref{eq:APx}~and~\ref{eq:APy}.
This source of systematic uncertainty was investigated through comparing the results with and without the free parameters
of the fitting functions on the asymmetry distributions. We obtained a systematic uncertainty $\sigma_{asymm}\sim 0$-$0.1$ for $A_{x}$
and $\sigma_{asymm}\sim 0$-$0.12$ for $A_{y}$.
The third source of the systematic uncertainty comes from the shape of the background.
This part of the systematic uncertainty consists of two components.
The first one is the difference between results of full and limited ranges of the background estimation,
see Fig.~\ref{fig:missmass}.
To do this, we keep the polynomial order to be the same but decrease the range of the background estimation.
The error from this part was estimated to be in the range of $\sigma_{bg1}\sim 0$-$0.04$ for both $A_{x}$ and $A_{y}$.
The other component of the error in the shape of the background was obtained using the difference between the
results of the background estimation using two different polynomial orders (6 and 8) but keeping the full
range.
The error from this source was estimated to be in the range of $\sigma_{bg2}\sim 0$-$0.04$.  
The total systematic uncertainty was obtained  by a quadratic sum of the four sources of systematic
errors, assuming them to be independent. The total systematic uncertainty varies between $0$-$0.16$ for $A_{x}$
and $0$-$0.2$ for $A_{y}$ depending on the configuration. We decided only to present data for which the total
systematic uncertainty is less than $0.08$ to have a meaningful comparison with the  calculations.
This operation only excluded  a  small fraction of the data points.

Figures~\ref{fig:Ax_45}-\ref{fig:Ay_134} show the results for $\theta_{2}=45^{\circ}$ and $134^{\circ}$ 
and different combinations of $\theta_{1}$ and $\phi_{12}$.
For some of the configurations where $\theta_{2}=45^{\circ}$, the data points
at $S>100$~MeV are not presented because we were not able to subtract
the background of the uncorrelated elastic break-up events unambiguously.
Also, for some $S$-values in which the peak-searching algorithm could not find
a peak around the expected missing-mass of neutron and therefore, we are not
presenting any data points.
For the case of $\theta_{2}=134^{\circ}$, the data points of the lower values
of $S$ ($S<40$~MeV) are not presented because the $S$-curve overlapped with the energy threshold of the
ball detectors and we could not separate the signal from the noise.
Note that our measured values for $A_{x}$ for coplanar configurations ($\phi_{12}=180^{\circ}$)
are compatible with zero. This is consistent with expectations based on parity conservation and,
therefore, shows that our analysis method is reliable. 
The present results correspond to the relatively large values of relative energy,
$E_{rel}>40$~MeV, between two final-state protons. Aside from the slight disagreements for the smaller
$\theta_1$ ($20^\circ$ and $25^\circ$) and $\phi_{12}$ ($140^\circ$), the results reveal a fairly good agreement between
data and the theoretical predictions for $A_{x}$ and $A_{y}$, confirming the previous report
on a part of the phase space of this experiment in which two protons scatter to the forward angles with
large azimuthal opening angle~\cite{MARDANPOUR2010}. It means that the theoretical models provide
a good description of analyzing powers for the cases with a large relative energy while they fail to
describe the data for lower values of relative energy in proton-deuteron break-up reaction~\cite{MARDANPOUR2010}.

For some configurations in which $\theta_{2}=134^{\circ}$, the two 3NF models provide different values of $A_{y}$ at the
intermediate values of $S$. Considering the statistical and systematic uncertainties,
our measurement cannot differentiate between these models. 
%
  
\section{Summary and conclusion}
\label{sec:summary}
In this article, a study of the vector analyzing powers in the proton-deuteron 
break-up reaction with an incident polarized proton-beam energy of $190$~MeV impinging on a
liquid deuterium target is presented.
The data were obtained exploiting the almost $4\pi$ detection system BINA.
The focus was on the part of the phase space at which one of the protons scattered
to intermediate and large polar angles.
The backward-ball of the BINA is used for the first time to extract break-up
observables in coincidence with the forward-wall.
We presented the vector analyzing powers,
$A_{x}$ and $A_{y}$, for a set of kinematical configurations in which one of
the protons scatters to the angles of $\theta_{2}=45^{\circ}$ and 
$\theta_{2}=134^{\circ}$ and the other to forward angles 
($12^{\circ}<\theta_{1}<32^{\circ}$).
This part of the break-up phase space at this beam energy is presented for the
first time.
At the large scattering angle of the second proton, $\theta_{2}=134^{\circ}$, the theoretical predictions
are in fair agreement with the data. However, the two 3NF models provide different values of analyzing 
powers at some kinematics. With the precision of the present experiment, one can, however, not distinguish
between these models.
The 3NF effects are expected to be small at the configurations with 
$\theta_{2}=45^{\circ}$ and at large azimuthal opening angles $\phi_{12}$.
For these configurations the theoretical predictions are also in
good agreement with the data. The results presented here mainly correspond
to larger values of relative energy between two final-state protons, and the
agreement between data and calculations are consistent with earlier measurements for this
range of relative energy. On the other hand, a significant disagreement
between data and theoretical predictions was reported at this energy for the configurations
in which the relative energy is smaller than $10$~MeV~\cite{MARDANPOUR2010}.
\textcolor{black}{The physics behind this discrepancy is not clear yet. A similar, but
less pronounced, discrepancy has been observed in the proton-deuteron break-up reaction
measured at a proton beam energy of $130$~MeV~\cite{Hajar2019}.}
Further investigation of all parts of the break-up phase space at different beam energies is needed to have a 
complete picture of nuclear forces and its dependence on different underlying dynamics.
\section{Acknowledgment}
We would like to express our gratitude to the AGOR accelerator group at KVI for
delivering the high-quality polarized beam.
This work was partly supported by the Polish National Science Center under
Grants No.~2012/05/B/ST-2/02556 and No.~2016/22/M/ST2/00173.
\bibliographystyle{h-physrev5}
\bibliography{Bibliography}

\end{document}